# P(VDF-TrFE)/BaTiO$_3$ nanoparticle composite films mediate piezoelectric stimulation and promote differentiation of SH-SY5Y neuroblastoma cells

*Giada Graziana Genchi\*, Luca Ceseracciu, Attilio Marino, Massimiliano Labardi, Sergio Marras, Francesca Pignatelli, Luca Bruschini, Virgilio Mattoli, Gianni Ciofani\**

Dr. G.G. Genchi, A. Marino, Dr. F. Pignatelli, Dr. V. Mattoli, Prof. G. Ciofani
Istituto Italiano di Tecnologia, Center for Micro-BioRobotics @SSSA, Viale Rinaldo Piaggio 34, 56025 Pontedera (Pisa), Italy
A. Marino
Scuola Superiore Sant'Anna, The BioRobotics Institute, Viale Rinaldo Piaggio 34, 56025 Pontedera (Pisa), Italy
Dr. L. Ceseracciu
Istituto Italiano di Tecnologia, Smart Materials, Nanophysics Department, Via Morego 30, 16163, Genova, Italy
Dr. S. Marras
Istituto Italiano di Tecnologia, Nanochemistry Department, Via Morego 30, 16163 Genova, Italy
Dr. M. Labardi
CNR-IPCF, Largo Pontecorvo 3, 56127 Pisa, Italy
Prof. L. Bruschini
University Hospital of Pisa, ENT Audiology and Phoniatry Unit, Via Paradisa 3, 56124 Pisa Italy
Prof. G. Ciofani
Politecnico di Torino, Department of Aerospace and Mechanical Engineering, Corso Duca degli Abruzzi 24, 10129 Torino, Italy

E-mail: giada.genchi@iit.it, gianni.ciofani@polito.it

**Abstract**

Poly(vinylidene fluoride-trifluoroethylene, P(VDF-TrFE)) and P(VDF-TrFE)/barium titanate nanoparticle (BTNP) films are prepared and tested as substrates for neuronal stimulation through direct piezoelectric effect. Films are characterized in terms of surface, mechanical, and piezoelectric features before *in vitro* testing on SH-SY5Y cells. In particular, BTNPs significantly improve piezoelectric properties of the films (4.5-fold increased $d_{31}$). Both kinds of films support good SH-SY5Y viability and differentiation. Ultrasound (US) stimulation is proven to elicit Ca$^{2+}$ transients and to enhance differentiation in cells grown on the piezoelectric substrates. For the first time in the literature, this study demonstrates the suitability of polymer/ceramic composite films and US for neuronal stimulation through direct piezoelectric effect.

1. Introduction

Electrical stimulation is used for the treatment of several conditions affecting the nervous and the musculoskeletal systems, like dopaminergic neuron loss in Parkinson's disease or bone loss after severe fracture or tumor surgery, thereby compensating altered electric communication among/to neurons and promoting tissue growth/regeneration.[1,2] Among the several methods recently proposed for the manipulation of electrical activity in cells and tissues (such as optogenetics[3] or microheating[4] of cells), those based on piezoelectric materials are highly promising concerning implementation in hardly accessible anatomical districts (as an example the cochlea) and activation thanks to simple environmental stimuli (mechanical waves such as sounds or ultrasounds, US) that can be achieved even in a wireless mode.

Piezoelectric materials and nanomaterials are increasingly being applied in the biomedical research for their interesting property of generating electric fields upon application of mechanical stresses. This property, also called "direct piezoelectric effect", enables the delivery of physiologically relevant electrical stimulation to a variety of cells and tissues in a wireless modality.[5] Our group performed several investigations on the interaction of piezoelectric nanomaterials and biological environments, as well as on the exploitation of their activity upon US stimulation. For instance, it was demonstrated that the simultaneous administration of boron nitride nanotubes and US to PC12 neuron-like cells determined the emission of significantly longer neurites (30% longer than the controls) by involvement of the TrkA receptor.[6] Very recently, we have also carefully evaluated the mechanisms of piezoelectric stimulation of cells in the presence of barium titanate nanoparticles (BaTiO$_3$ NPs, BNTPs) and of US in SH-SY5Y neuron-like cells.[7]

In the present work, we explore the possibility to use composite films made of a piezoelectric co-polymer (poly(vinylidene fluoride-trifluoroethylene, P(VDF-TrFE)) and piezoelectric ceramic nanoparticles (again, BTNPs) to stimulate SH-SY5Y cells, a human neuroblastoma cell line chosen as neuronal model, through exposure to US. P(VDF-TrFE) is well known for its good piezoelectric, mechanical and processability properties,[8] whereas BTNPs offer superior piezoelectric performances that can further improve those of P(VDF-TrFE).[9] Cell responses under proliferative and differentiative conditions were investigated, in particular by applying an acute and a chronic US stimulation during cell differentiation. Our results provide the first evidences of US-activated piezoelectricity of P(VDF-TrFE)/BTNP composite films, useful for wireless neuronal stimulation.

2. Results

2.1. Characterization of the surface of P(VDF-TrFE) and P(VDF-TrFE)/BTNP films

Scanning electron microscopy (SEM) images of plain and BTNP doped films are reported in **Figure 1a** and **1b**, respectively. **Figure 1b** shows the presence of homogeneously dispersed nanoparticles (visible as brighter spots) in composite samples. SEM images of the fracture surface of cryofractured films are reported as **Figure 1c** (plain) and **1d** (composite), the latter clearly showing that BTNPs are homogeneously distributed in the doped substrates. **Figure 1e** (plain) and **1f** (composite) represent the fracture surface of cryofractured samples at higher magnification. In **Figure 1e**, the fracture surface of the polymeric sample shows a peculiar pattern, probably due to the different fracture behaviour of amorphous and crystalline phases. The presence of BTNPs in the composite sample (**Figure 1f**) induces, on the other hand, the formation of cavities surrounding particles, which we attribute to differential shrinkage and debonding from the thermal expansion mismatch (approximately $7 \cdot 10^{-5}$ K for P(VDF-TrFE)[10] and $2 \cdot 10^{-5}$ K for P(VDF-TrFE)/BaTiO$_3$[11]).

Atomic force microscopy (AFM) images of the surfaces of the films are reported as **Figure 2a** and **2b** (again plain and composite, respectively). Root mean square (RMS) roughness quantification was performed analyzing these AFM images and reported in **Table 1**. These data demonstrate that the P(VDF-TrFE)/BTNP films present a much higher surface roughness (~212 nm) with respect to the plain P(VDF-TrFE) film (~63 nm).

Piezoresponse force microscopy (PFM) maps of small regions of the film surface are reported as **Figure 2c** and **2d** (once again, plain and composite, respectively). In both figures, the brightness of the piezoresponse map corresponds to the piezoelectric displacement of the film surface under the application of a given electric field concentrated in a nanometric region close to the scanning probe tip. In **Figure 2d**, the background piezoresponse is due to the plain P(VDF-TrFE), while the enhanced piezoresponse amplitude corresponds to the BTNPs. Quantitative estimation of the local piezoelectric coefficient $d_{33}$ is indicated (pm/V) by the scale bar beside **Figure 2c** and **2d**. The scale provides values that are not directly comparable to the ones measured by piezoelectric strain measurement, but just give an indication of the enhanced piezoactivity due to the presence of the BTNPs.

**Table 1** reports in detail the results of the piezoelectric property measurements. The converse piezoelectric coefficient $d_{31}$ was 11.8 pm/V for P(VDF-TrFE) and 53.5 pm/V for P(VDF-TrFE)/BTNP films (4.5-fold increment), whereas the direct piezoelectric coefficient $g_{31}$ was 0.11 for P(VDF-TrFE) and 0.24 for P(VDF-TrFE)/BTNP films (2.2-fold increment). Polytetrafluoroethylene (PTFE) was used as control non-piezoelectric material.

2.2. Characterization of the bulk properties of P(VDF-TrFE) and P(VDF-TrFE)/BTNP films

Differential scanning calorimetry (DSC) thermograms of the two typologies of films are reported in **Figure 3a**. The plain films present three endothermal peaks: one sharp peak is visible at ~150°C, corresponding to the melting temperature of the crystalline component of the co-polymer, and two shallow peaks are visible at ~105°C and ~96°C, corresponding to the Curie temperatures of the different crystalline phases of the co-polymer. The composite P(VDF-TrFE)/BTNP film presents the same peaks at ~150°C and ~105°C as P(VDF-TrFE), whereas that one ~96°C is flattened. A very small peak at ~127°C is also present corresponding to the Curie temperature of barium titanate.

Thermograms obtained with thermogravimetric analysis (**Figure 3b**) demonstrated that, as expected, nanoparticles represent about the 60% (w/w) of the composite films (see **Table 2** for details).

X-ray diffraction (XRD) patterns (**Figure 3c**) exhibit typical peaks of the α phase at 17.14° and 40.12°, and of the β phase at 19.98°, 35.13° and 40.77°. The amorphous phase is characterized by a broad halo at 18.73°. Based on values from the literature,[12,13] fitting of the patterns enabled the quantification of the α and β phases, and thus of the overall crystallinity of the samples. The percentage of the α phase in the P(VDF-TrFE) films was ~30%, whereas that one of the P(VDF-TrFE)/BTNP films was ~15 % (**Table 2**). The percentage of the β phase in the P(DVF-TrFE) films was ~30%, whereas it was ~50% in the P(VDF-TrFE)/BTNP films. The typical peaks of tetragonal BTNPs in composite films are shown in Supporting Information, **Figure S1**. Detailed XRD results are reported in **Table 2**.

Representative stress-strain curves of the films are shown in **Figure 4a**. The behavior of the P(VDF-TrFE) films is typical of a ductile material, with a clear yield corresponding to the formation of a plastic neck, and with a long plateau corresponding to the extension of the plastic zone along the sample length until failure. The behavior of P(VDF-TrFE)/BTNP films is very different, being characterized by: 1) higher Young's modulus due to the BTNPs; 2) fracture taking place at lower values, and 3) larger scatter (indicating brittleness of the composite) in comparison to the plain P(VDF-TrFE) films. Stress/strain curves were used to quantify Young's modulus (*E*, **Figure 4b**), ultimate tensile strength (*UTS*, **Figure 4c**), extension at maximum load (*EML*, **Figure 4d**), and extension at break (*EB*, **Figure 4e**) for the two substrates. In particular, *E* was found 404 ± 40 MPa for P(VDF-TrFE) and 784 ± 103 MPa for P(VDF-TrFE)/BTNP films. All other data are reported in **Table 2**. With the exception of the *UTS*, all of the parameters were significantly different between P(VDF-TrFE) and P(VDF-TrFE)/BTNP films ($p < 0.05$).

2.3. SH-SY5Y cell response to films under proliferative conditions

SH-SY5Y cell viability was assessed with Live/Dead staining, and resulted high and comparable on all of the substrates after 24 h from seeding. As shown in **Figure 5a**, only live cells (stained in green) could be imaged with fluorescence microscopy, while no dead cells (expected to appear in red) could be detected. However, we have to highlight as the PicoGreen assay (**Figure 5b**) showed a decreased proliferation of cells cultured on the P(VDF-TrFE) and P(VDF-TrFE)/BTNP films with respect to the control substrate (a standard Ibidi film).

2.4. SH-SY5Y cell response to films under differentiative conditions

Variations of the fluorescence signal *(ΔF)* due to $Ca^{2+}$ transients normalized to the background fluorescence ($F_0$) are plotted for all of the cultures that underwent an acute US stimulation in **Figure 6**. On the left, time courses of the *ΔF/F$_0$* traces are reported with insets representing the same traces only in the 20-40 s timeframe to show the sharp increase of the fluorescence signal at the application of US. On the right, a representative calcium imaging time-lapse frame (at $t = 50$ s). The application of the acute stimulation protocol demonstrated that only piezoelectric substrates were able to evoke $Ca^{2+}$ transients. The amplitude of these peaks was *ΔF/F$_0$* = 3.54 ± 0.35 on P(VDF-TrFE) films and *ΔF/F$_0$* = 7.16 ± 0.51 on P(VDF-TrFE)/BTNP films, the latter being significantly ($p < 0.05$) higher with respect to that observed in cells stimulated on plain substrates.

**Figure 7a** reports images of SH-SY5Y cell cultures (with β3-tubulin fluorescently stained in green, and nuclei in blue) after 6 days of differentiation on the different substrates, with and without chronic US stimulation. Neurites are qualitatively longer on the P(VDF-TrFE) and on the P(VDF-TrFE)/BTNP films after application of US.

The number of β3-tubulin positive cells over the total cell number and the neurite lengths were quantified with ImageJ analysis on confocal images. The number of β3-tubulin positive cells/total cell number was expressed as percentage ± standard deviation (**Figure 7b**), while neurite lengths were reported as median values with their interval of confidence at 95% (**Figure 7c**). In the absence of US stimulation, the β3-tubulin positive cells were 18 ± 3 %, 17 ± 3 % and 19 ± 2 % on Ibidi (as control), P(VDF-TrFE), and P(VDF-TrFE)/BTNP films, respectively. In the presence of US stimulation, the β3-tubulin positive cells were 18 ± 3 %, 43 ± 6 % and 79 ± 6 % on Ibidi (as control), P(VDF-TrFE), and P(VDF-TrFE)/BTNP films, respectively. These results demonstrate that the percentage of β3-tubulin positive cells was significantly enhanced on piezoelectric substrates in comparison to the controls, but only in the presence of US chronic stimulation. Coherent results were found for neurite length quantification. In the absence of US stimulation, median neurite lengths were 31 µm, 32 µm and 29 µm in cells cultured respectively on Ibidi (as control), P(VDF-TrFE), and P(VDF-TrFE)/BTNP films. In the presence of US stimulation, median neurite lengths were 28 µm, 40 µm and 49 µm in cells cultured respectively on Ibidi (as control), P(VDF-TrFE), and P(VDF-TrFE)/BTNP films. These results demonstrate that neurite elongation was significantly increased on piezoelectric substrates in comparison to the controls, but only if US had been applied. Most importantly, the presence of BNTPs in the P(VDF-TRFE)/BNTP films determines the emission of significantly longer neurites with respect to the plain P(VDF-TrFE) films, thus confirming an enhanced piezoelectric activity in the composite films. SEM images of SH-SY5Y cells after 6 days of differentiation on plain and BTNP doped films (and on Ibidi film as control) are reported in **Figure 8**. These images show the close interaction of differentiated cells with the surface of all of the substrates, and the presence of evidently longer neurites in cells on the piezoelectric substrates after US stimulation, in agreement with the previously reported quantitative analysis.

3. Discussion

In the present work, P(VDF-TrFE) and P(VDF-TrFE)/BTNP films were prepared with an easy sonication and cast/annealing method. SEM imaging of sample surface and of the fracture surface of cryofractured substrates demonstrated high homogeneity of the films, in particular in terms of nanoparticle distribution. This feature is extremely important, since nanoparticle distribution represents one of the several factors that influence piezoelectric response of composite films and has been demonstrated to improve piezoelectric output in nanogenerators.[9,14]

AFM demonstrated that the surface roughness underwent a ~3-fold increase upon BTNP dispersion in P(VDF-TrFE). Surface roughness is well known to modulate cells responses,[15-17] however the different roughness of P(VDF-TrFE) and P(VDF-TrFE)/BTNP films did not affect SH-SY5Y differentiative behavior, as demonstrated by the comparable percentage of β3-tubulin positive cells and neurite lengths at the end of a 6-day differentiation without ultrasound stimulation.

For P(VDF-TrFE) and P(VDF-TrFE)/BTNP film preparation, a low annealing temperature was chosen as a compromise between the need of increasing the ferroelectric β phase of P(VDF-TrFE)[18] and the maintenance of a satisfactory deformability despite the high amount of the ceramic component in the composite films.[19]

DSC thermal analysis of the films demonstrated that both films had very similar enthalpies of fusion. The melting peak of the P(VDF-TrFE)/BTNP film was slightly sharper and lower than that of the P(VDF-TrFE) film. Two peaks were also found that can be associated to the Curie transitions from the polar ferroelectric to the non-polar paraelectric phases in the copolymer.

The origin of these two transitions was discussed by several authors,[20,21] and is generally ascribed to a mixture of two crystalline polar ferroelectric β phases in the copolymer. In particular, the higher temperature peak is more intense in the P(VDF-TrFE)/BTNP films than in the P(VDF-TrFE) films, and can be attributed to a more ordered β phase with larger domains. Furthermore, the enthalpy of fusion of the Curie transition increased from 22 to 28 J/g by addition of BTNPs, pointing out the presence of larger crystalline ferroelectric β phases in the composite films. These results are in line with previous reports on the nucleation of the β phase induced by the BTNPs in PVDF.[22] Finally, the thermogram of the P(VDF-TrFE)/BTNP films shows an additional little endothermic peak at 126°C that can be attributed to the Curie transition of barium titanate.[23]

XRD analysis proved an increased crystallinity in the composite films, by revealing that this increment is indeed due to a higher β phase percentage induced by BTNPs.

Extensometry analysis confirmed that the dispersion of BTNPs in P(VDF-TrFE) dramatically changed the mechanical properties of the composite films with respect to the plain polymeric films. The observed decrease in *UTS* and elongation of the composite material is quite a common phenomenon in reinforced plastics. It arises from the fact that hard particles act as defects in the matrix, so that plastic deformation is reached for lower stress than the plain counterpart, and the plastic/viscous flow of the polymer chains is hindered. Although both *UTS* and elongation are lower than those of plain P(VDF-TrFE), they are well above the ones expected during service, coming both from sample handling and piezoelectric stimulation. A useful parameter to evaluate the resistance to handling is the radius of curvature that the sample can withstand without failing: the thickness (*t*) and the extension at maximum load (*EML*) of our sample indeed allow for a radius of curvature (*R*) of about 6 mm, calculated from the following equation:

$$R = \frac{t}{2EML} \qquad (1)$$

In order to enlighten the matrix-filler relationship, the measured values of Young's modulus were compared to models from the literature, also considering the microscopy observations of the respective microstructures. The increase in Young's modulus ($E_c$) of the P(VDF-TrFE)/BTNP films can be well described by the Ogorkiewicz and Weidmann's cube-within-cube model:[24]

$$E_c = E_P \left[ 1 + \frac{2V_{BT}^{2/3}}{(mV_{BT}^{1/3} + 1)/(m-1) - V_{BT}^{2/3}} \right] \quad (2)$$

where the subscripts $P$ and $BT$ respectively refer to the polymeric and the nanoparticle phase, $m = E_{BT}/E_P$ (with $E_{BT}$ = 70 GPa), and $V_{BT}$ = 0.17 is the relative volume of the particles, calculated through the densities $\rho_P$ = 1.78 g/cm$^3$ and $\rho_{BT}$ = 6.02 g/cm$^3$. The mechanical response of the composite is thus well represented by a model that is generally considered as indicative of a good matrix-reinforcement interface.[25] An additional indication of good matrix-filler interface can be found in the value of *UTS*, which is not significantly lower, albeit more scattered, than that one of the P(VDF-TrFE) film.

Environmental stiffness is well known to affect neuron responses,[26] however we found comparable neurite lengths in SH-SY5Y cells differentiated on all of the substrates in the absence of ultrasound stimulation. Although preliminary, these SH-SY5Y differentiation results suggest that substrate roughness and stiffness may play minor roles in addressing cell response compared to piezoelectricity, but of course further studies are necessary to clarify these aspects.

The culture of SH-SY5Y cells under proliferative conditions demonstrated high and comparable viability on all of the substrates after 24 h from seeding, despite a small, yet significant, decrement of cell proliferation after 72 h on the piezoelectric substrates. These data are however in line with the literature. For instance, a recent study demonstrated that microporous P(VDF-TrFE) membranes prepared by casting and solvent evaporation supported C2C12 myoblast and MC3T3-E1 pre-osteblast adhesion; however, C2C12 and MC3T3-E1 proliferation was found to be decreased in comparison to control substrates after 48 h and 72 h of culture.[27] Another study showed that cell viability of human alveolar osteoblasts on P(VDF-TrFE)/BTNP films was high, well-retained and comparable to controls (expanded polytetrafluoroethylene, ePTFE films) for a 10 day period of culture. Osteoblast proliferation was instead significantly decreased on P(VDF-TrFE)/BTNP films compared to controls. Alkaline phosphatase activity was higher on P(VDF-TrFE)/BTNP films in comparison to ePTFE films after 7 days of culture; moreover, bone-like nodule formation exhibiting osteocyte lacunae occurred only on composite films, suggesting improved cell differentiation on P(VDF-TrFE)/BTNP films compared to the controls.[28]

In our study, SH-SY5Y were cultured under differentiative conditions and exposed to an acute ultrasound stimulation on the different substrates at the end of a 6-day differentiation period. $Ca^{2+}$ transients were observed only on piezoelectric films upon stimulation. The amplitude of these transients was significantly higher in cells cultured on the composite films, indicating better piezoelectric transduction operated by the P(VDF-TrFE)/BTNP substrates compared to the plain polymeric films. These results are in line with what we found concerning a 24 h interaction of gum Arabic-coated BTNPs with SH-SY5Y cells differentiated for 4 days. By mostly interacting with cell membranes, BTNPs induced higher amplitude $Ca^{2+}$ transients in SH-SY5Y cells exposed to ultrasounds compared to the simple ultrasound stimulation. Interestingly, the administration of inhibitors collectively proved that opening of both $Ca^{2+}$ and $Na^+$ voltage-gated channels on the cellular membranes was at the base of the observation of high amplitude $Ca^{2+}$ transients. Further, we calculated that a local voltage of ~2 mV is generated by BTNP clusters when exposed to ultrasounds, which very likely determines the opening of voltage-gated channels on cell membranes.[7]

SH-SY5Y cells were also cultured under differentiative conditions and exposed to a chronic ultrasound stimulation on the different substrates for the whole differentiation period. β3-tubulin was used as a typical marker of differentiation in neuronal cells.[29] With the analysis of confocal microscopy images, differentiation (evaluated in terms of percentage of β3-tubulin positive cells and of neurite length) was found to be enhanced on piezoelectric substrates after ultrasound stimulation. Interestingly, from the literature it is known that the expression of this marker in SH-SY5Y cells increases with culture time.[30] Our data of percentage of β3-tubulin positive cells demonstrate that the differentiation is higher on piezoelectric substrates after US stimulation. Neurite length was moreover significantly improved on P(VDF-TrFE)/BTNP films, again denoting an improved piezoelectric transduction due to the dispersion of BTNPs in the copolymer.

In the literature, other examples can be found describing enhanced neuronal differentiation on P(VDF-TrFE) films. For instance, P(VDF-TrFE) was used for the culture of human neural stem/progenitor cells, and β3-tubulin positive cells could be found only on annealed films after 9 days of culture. P(VDF-TrFE) was also processed into micro- and nanofibers of different orientation and it was coated with laminin. In this case, longer neurites (∼50 μm) were emitted by cells on aligned annealed microfibers compared to non-annealed microfibers.[17] Annealed aligned fibers of P(VDF-TrFE) also promoted the extension of longer neurites (∼2 mm) in dorsal root ganglion cells with respect to annealed random fibers.[31]

In another study, stretched and poled PVDF films were tested for spinal cord neuron culture in the presence and in the absence of mechanical vibrations. After 4 days of culture on PVDF films, neurite outgrowth was increased by ∼80% in cells exposed to vibrations compared to those not exposed to vibrations. Neurite branching was increased as well. These parameters were not affected when cells were cultured on non-piezoelectric PVDF films, thereby excluding neuritogenesis improvement due only to vibrations.[32]

By comparing the present data with the literature, our results can be considered the first evidences on wireless electrical stimulation of neuronal cells thanks to piezoelectric transduction mediated by polymer-ceramic composite films, that thus represent a promising

"smart" platform for a wide range of tissue engineering and regenerative medicine applications.

The developed piezoelectric films can in fact be envisaged as active scaffolds able to provide electric stimulation to cell cultures, including muscle,[33] heart,[34] bone,[35] and stem cells.[36] Given the high biocompatibility of the exploited materials, we deem that this approach could find soon translational applications, in particular in those cases where a "contactless" electrical stimulation is highly desirable. Neuronal interfaces based on P(VDF-TrFE)/BaTiO$_3$ nanoparticle composite might be exploited, as an example, to drive efficient physical cues to the nerves during the regeneration process[37] or as functional bio-hybrids for sensing restoring.[38]

Overall, the developed films can act as active "patches" to elicit appropriate responses in electrically-sensitive tissues, through a wireless technique that minimizes the invasiveness of traditional *in vivo* electric stimulation.

4. Conclusion

This study showed that the dispersion of tetragonal piezoelectric BaTiO$_3$ nanoparticles into a P(VDF-TrFE) solution enabled the achievement of composite films characterized by high homogeneity, improved crystallinity and notably enhanced piezoelectric properties. Piezoelectric films supported SH-SY5Y neuroblastoma cell cultures under either proliferative and differentiative conditions. When differentiated cultures were exposed to ultrasound stimulation, Ca$^{2+}$ transients were demonstrated to occur on both P(VDF-TrFE) and P(VDF-TrFE)/BTNP films as evidence of neuronal stimulation due to piezoelectric effect. Amplitude of Ca$^{2+}$ transients was affected by film composition, resulting to be significantly higher on the composites, by virtue of the improved piezoelectric features. When differentiating cultures were exposed to chronic ultrasound stimulation, neuronal maturation was demonstrated to be enhanced on piezoelectric substrates. Most importantly, neurite length was found to be significantly improved on P(VDF-TrFE)/BTNP films upon ultrasound stimulation. These results are highly promising concerning the potentiality of P(VDF-TrFE)/BTNP films as devices for neuron stimulation, indicated for those conditions requiring long-term treatments, as for instance auditory deficits due to cochlear hairy cell alterations.[1]

5. Experimental Section

*5.1. P(VDF-TrFE) and P(VDF-TrFE)/BTNP film preparation*

For film preparation, P(VDF-TrFE) from Piezotech (70/30% mol copolymer) and barium titanate nanoparticles (BTNPs) from NanoAmor (tetragonal, ~300 nm) were used. For plain films, 1 g of P(VDF-TrFE) was dissolved in 10 ml of methylethylketone (MEK, Carlo Erba) with the aid of a Sonidel tip sonicator set at 8 W for 10 min. For composite films, 400 mg of P(VDF-TrFE) and 600 mg of BTNPs were mixed and then 10 ml of MEK were added. The 40% P(VDF-TrFE)/60% BTNP (w/w) mixture was homogenized through sonication at 8 W for 10 min. The mixtures were allowed to rest for 30 min prior to further processing. Films were prepared by casting 800 μl of both mixtures on Ibidi film pieces (2.5 cm x 2.5 cm) and annealing at 40°C on a hot plate for 4 h. The residual solvent was removed by placing samples under vacuum for 12 h. Samples were exposed to $O_2$ plasma (25 sccm, 50 W, 120 s, 0.5 mbar) to increase surface hydrophilicity, and then they were characterized as specified in the following. Prior to cell culture, samples were cut into 7 mm x 7 mm pieces and sterilized by incubation with penicillin/streptomycin (100 U/ml) in phosphate buffered saline (PBS) for 30 min. For calcium imaging, films were cut into 35 mm diameter disks.

*5.2. Characterization of surface and piezoelectricity of P(VDF-TrFE) and P(VDF-TrFE)/BTNP films*

Surface morphology of the films was characterized with a FEI Helios 600i scanning electron microscope by applying a 10 kV acceleration and a 86 pA current. Prior to imaging, samples were gold-sputtered (25 mA, 60 s, coating thickness ~10 nm).

Nanoparticle dispersion homogeneity within the films was investigated with a JEOL JSM-7500F scanning electron microscope equipped with a cold FEG working at 5 kV. Prior to imaging, the films were immersed into liquid nitrogen and then bended to obtain a clean fracture surface. Samples were carbon-coated before imaging (evaporation time: 2800 ms *per* pulse; 3 pulses; coating thickness: ~15 nm).

Surface topography was mapped with a Veeco Instruments Multimode atomic force microscope equipped with Nanoscope IIIa controller. For tapping mode atomic force microscopy (AFM), non-contact mode silicon cantilevers were used (Veeco RTESP10, with 265-311 kHz nominal resonant frequency, and 20-80 N/m nominal spring constant). The same instrument was operated as a piezoresponse force microscope (PFM) in the contact-resonance mode[39], by using contact mode silicon cantilevers (NanoWorld Arrow-cont, with 14 kHz nominal resonant frequency, and 0.2 N/m nominal spring constant).

Piezoelectric transduction was measured after cutting the films into stripes of size 20 mm x 5 mm x 0.08 mm (*xyz*), and loading them into a home-made setup briefly described in the following. One end of the sample was fixed to a holder, attached to a harmonic steel slab spring by means of Teflon clamps. The other end of the sample was attached to a small microtranslator, enabling pre-tension of the sample by Teflon clamps as well. An electric potential could be applied in transverse direction (*z*) through a pair of metal slabs, positioned very close to the stripe wide surface (*xy*). In this way, the transverse electric field produced a longitudinal deformation (along *x*) with magnitude proportional to the piezoelectric coefficient $d_{31}$ of the material. Alternating electrical drive at the resonance frequency of the spring produced an enhancement of its vibration by a factor of about 20, due to the harmonic oscillator quality factor *Q*. Bending of the steel slab was detected by the optical lever method, where a laser beam was reflected by a mirror attached on the spring and its deflection was detected by a four-quadrant split photodiode. The setup is schematically depicted in Supporting Information, **Figure S2**.

The converse piezoelectric coefficient $d_{31}$ was calculated, in the used setup configuration, as:

$$d_{31} = 2.1 \cdot 10^4 \frac{D\varepsilon_s}{l_e} \frac{V_{out,RMS}}{Q} \tag{3}$$

where $D$ is the spacing between electrodes, $l_e$ the electrode size along $x$, $V_{out,RMS}$ the vibration amplitude output of the apparatus, $Q$ the oscillator quality factor, and $\varepsilon_S = \varepsilon_d - \frac{D_d}{D}(\varepsilon_d - 1)$, where $\varepsilon_d$ is the effective dielectric constant of the composite and $D_d$ the film thickness. $\varepsilon_d$ was calculated with the model from Bhimasankaram and co-workers (BSP)[40] as:

$$\varepsilon_d = \frac{\varepsilon_P(1-q) + \varepsilon_{BT}q[3\varepsilon_P/(\varepsilon_{BT}+2\varepsilon_P)][1+3q(\varepsilon_{BT}-\varepsilon_P)/(\varepsilon_{BT}+2\varepsilon_P)]}{(1-q) + q[3\varepsilon_P/(\varepsilon_{BT}+2\varepsilon_P)][1+3q(\varepsilon_{BT}-\varepsilon_P)/(\varepsilon_{BT}+2\varepsilon_P)]} \tag{4}$$

with $\varepsilon_P$ the dielectric constant of the polymer, $\varepsilon_{BT}$ that one of the BTNPs, and $q$ the volume fraction of the BTNPs. Values used in the present experiment were $D$ = 1 mm, $l_e$ = 6.5 mm, $D_d$ = 0.08 mm, $\varepsilon_P$ = 12, $\varepsilon_{BT}$ = 1250, $\rho_P$ = 1.78 g/cm$^3$, $\rho_{BT}$ = 6.02 g/cm$^3$. The value of the calibration factor in (3) depends on various characteristic of the setup in the used configuration, as optical lever length, photodiode sensitivity, and so on.

The direct piezoelectric coefficient $g_{31}$ was calculated as $g_{31} = \frac{d_{31}}{\varepsilon_0 \varepsilon_d}$, being $\varepsilon_0$ = 8.82·10$^{-12}$ F/m the dielectric constant of vacuum.

*5.3. Characterization of P(VDF-TrFE) and P(VDF-TrFE)/BTNP film bulk properties*

Bulk properties of the films were investigated with differential scanning calorimetry (DSC), thermogravimetric analysis (TGA), X-ray diffraction (XRD) analysis and extensometry. DSC was performed with a Mettler Toledo STARe system equipped with a DSC-1 calorimetric cell. For the analysis, a small piece (~5 mg) of film was put in standard aluminum pans (Mettler Toledo), and two consecutive scans were performed over a 20-200°C temperature range with a 10°C/min heating rate. Analysis was conducted on the second thermal scan.[21]

TGA was performed on small pieces of films (~5 mg) with a TA Instruments Q500 TA apparatus. The scans were performed in the 30-800°C temperature range by using a 50 ml/min nitrogen flow, a 10 °C/min heating rate, and platinum pans.

XRD patterns were recorded on a Rigaku SmartLab X-Ray diffractometer equipped with a 9kW CuKα rotating anode, operating at 40 kV and 150 mA. A Göbel mirror was used to convert the divergent X-ray beam into a parallel beam and to suppress the Cu Kβ radiation. Diffraction patterns were collected at room temperature over a 4-50° angular range, and with a step size of 0.05°. XRD data analysis was carried out using PDXL 2.1 software from Rigaku.

Extensometry was performed with an Instron 3365 dual column universal testing machine. For uniaxial tensile tests, samples were cut into 20 mm x 1.5 mm stripes and then loaded on the clamps. Sample thickness was measured prior to tensile test with a Mitutoyo digital micrometer, and a thickness of ~50 μm and ~120 μm was respectively found for P(VDF-TrFE) and for P(VDF-TrFE)/BTNP films. Displacement was applied with the constant rate of 2 mm/min. From the stress-strain curves, Young's modulus, yield stress and strain, extension at maximum load and extension at break were extracted.

### 5.4. SH-SY5Y cell culture on P(VDF-TrFE) and P(VDF-TrFE)/BTNP films

SH-SY5Y human neuroblastoma cells (ATCC® CRL-2266™) were used as a neuronal model. SH-SY5Y were seeded at a density of 10,000 cells/cm$^2$ over the films cut into 7 mm x 7 mm pieces, and placed into 24-well plates. Ibidi film pieces (Ibidi) were used as non-piezoelectric controls. The Ibidi film is a cyclic olefin copolymer (COC) with demonstrated biocompatibility (conformity to the EN ISO 10993-5 standard). This polymeric film was chosen as control substrate as it can easily be cut into pieces of desired shape and size, and enables high resolution imagining thanks to its refraction index identical to that of glass. It is widely used in the literature.[41] Cells were cultured with DMEM-F12 medium (Gibco), added with 10% fetal bovine serum (FBS, Gibco), penicillin/streptomycin (100 U/ml, Gibco) and amphotericin B (2.5 μg/ml, Sigma Aldrich) for proliferation studies. Cells were cultured with DMEM, added with retinoic acid (10 μM, Sigma Aldrich), 1% FBS, penicillin/streptomycin and amphotericin B for differentiation studies. Incubation was performed at 37°C in a 5% $CO_2$ humidity-saturated atmosphere. Cell culture medium was changed every two days.

*5.5. SH-SY5Y cell response under proliferative conditions to P(VDF-TrFE) and P(VDF-TrFE)/BTNP films*

Cell response to the substrates under proliferative conditions was investigated in terms of viability with Live/Dead staining (Invitrogen) after 24 h from seeding, and of ds-DNA quantity with PicoGreen assay (Invitrogen) after 24 h and 72 h from seeding.

Live/Dead staining was performed following the manufacturer's instructions. Briefly, SH-SY5Y cells were incubated with cell culture medium added with calcein AM (2 μM), EthD-1 (4 μM), and Hoechst 33342 (1 μg/ml, for nucleus counterstaining) at 37°C in the dark for 10 min, then they were washed with phosphate buffered saline (PBS, 1X with calcium and magnesium) and immediately imaged with an inverted epifluorescence microscope (Eclipse Ti, Nikon).

PicoGreen assay was performed following the manufacturer's instructions. Briefly, samples were washed with 1X PBS (without calcium and magnesium), and lysed with 500 μl of ultrapure water and three cycles of freezing/thawing. Then, 100 μl of working solution were mixed to 50 μl of cell lysate and to 150 μl of solution containing the PicoGreen dye. Samples were incubated in the dark for 10 min, and finally fluorescence was read with a microplate reader (Victor X3, Perkin Elmer) at an emission wavelength of 535 nm by using a 485 nm excitation wavelength. To convert fluorescence into cell number, a calibration curve was drawn by processing SH-SY5Y pellets of known increasing cell number (1000, 5000, 10000, 50000, and 100000 cells), and assaying them as described above.

*5.6. SH-SY5Y cell response under differentiative conditions to P(VDF-TrFE) and P(VDF-TrFE)/BTNP films*

Cell response to the substrates under differentiative conditions was investigated by applying stimulation with ultrasounds (US) with a SonoPore KTAC-4000 sonoporator set at 1 W/cm$^2$ (100 Hz burst rate) and equipped with a KP-S20 probe.

In order to perform calcium imaging experiments, US were applied for 40 s to SH-SY5Y cultures. This stimulation protocol was chosen for imaging Ca$^{2+}$ transients in SH-SY5Y cells and thus for providing a direct evidence of piezoelectric transduction mediated by P(VDF-TrFE) and P(VDF-TrFE)/BTNP films. Prior to stimulation and imaging, cells were differentiated for 6 days on films positioned in Ibidi μ-dishes (35 mm). For imaging, SH-SY5Y cells were rinsed with DMEM (serum and phenol red free) and incubated with Fluo-4 AM (1 μM, Invitrogen) for 30 min at 37°C. Then, they were again rinsed and incubated with artificial cerebrospinal fluid (ACSF, composed by 140 mM NaCl, 5 mM KCl, 2 mM CaCl$_2$, 2 mM MgCl$_2$, 10 mM HEPES, and 10 mM d-glucose at pH=7.4; all reagents from Sigma Aldrich). For confocal observation (C2s, Nikon), the films were inverted, fixed in Ibidi dishes with the aid a home-made silicone ring (inner diameter 23 mm, outer diameter 35 mm), and fully immersed in ACSF. The amplitude of the Ca$^{2+}$ transients was obtained as specified elsewhere.[7]

Differentiating cells were also "chronically" stimulated, by applying US for 5 s twice a day, for 6 days. This protocol was chosen to demonstrate that the piezoelectric transduction mediated by P(VDF-TrFE) and P(VDF-TrFE)/BTNP films could improve development of neuronal phenotype in our cell model.

*5.7. SH-SY5Y immunostaining and confocal image analysis*

Soon after the last US application in the chronic stimulation protocol, SH-SY5Y cells were washed with PBS (with calcium and magnesium) and fixed with paraformaldehyde (4% in PBS, from Sigma Aldrich) for 18 min at 4°C. Then, cells were again rinsed with PBS and permeabilized with Triton X100 (0.1% in PBS, from Sigma Aldrich). Aspecific binding sites were saturated with diluted goat serum (GS, 10% in PBS; Thermo Scientific) for 1 h at 37°C. Then, a primary antibody (rabbit polyclonal IgG, Sigma Aldrich) against β3-tubulin was diluted 1:75 in 10% GS, and the solution was provided to fixed cells for 1 h at 37°C. Samples were then rinsed four times with 10% GS (5 min each rinse), thereafter a secondary antibody (goat polyclonal IgG, Thermo Scientific) was diluted 1:200 in 10% GS, and this solution (also containing 1 μM DAPI for nucleus counterstaining) was provided to cells for 45 min at 37°C. After one rinse with high salt PBS (0.45 M NaCl, 1 min) and with PBS, samples underwent confocal imaging. The efficiency of neural differentiation was evaluated by measuring both the percentage of the cells immunopositive for the β3-tubulin marker and the neurite lengths thanks to the Multi Measure plug-in of ImageJ software (http://rsbweb.nih.gov/ij/). Over 1000 cells and 400 neurites *per* substrate typology were considered for statistical analysis.

5.8. *SH-SY5Y imaging with SEM*

Cell-material interaction and cellular protrusions after 6 days of differentiation were qualitatively investigated by SEM imaging. Prior to imaging, samples were fixed with 4% PFA in PBS for 30 min, and with 5% glutaraldehyde in water for 4 h. Samples were then dehydrated with an ethanol gradient (25%, 50%, 75% and 100% ethanol in water, 15 min each step) and gold-sputtered (25 mA, 60 s, coating thickness ~10 nm) before observation.

*5.9. Statistical analysis*

All studies were carried out in triplicate. Extensometry and cell density quantification data are expressed as mean ± standard deviation. Calcium transients are expressed as $\Delta F/F_0$ peak mean ± standard error. The number of β3-tubulin positive cells/total cell number are expressed as percentage ± standard deviation. Neurite lengths are expressed as median ± 95% confidence interval in box-plots. Extensometry data were analyzed with one-way analysis of variance. Biological data were analyzed with Kruskal-Wallis nonparametric rank sum test, followed by Nemenyi-Damico-Wolfe-Dunn *post-hoc* test by using *R* software (http://www.r-project.org/). Data with *p* value < 0.05 were considered statistically significant.

**Supporting Information**
Supporting Information is available from the Wiley Online Library or from the author.

**Acknowledgements**
Alice Scarpellini (Nanochemistry Department, Istituto Italiano di Tecnologia) and Lara Marini (Nanophysics Department, Istituto Italiano di Tecnologia) are gratefully acknowledged for their assistance in SEM and TGA. Mauro Lucchesi (CNR-IPCF) is moreover acknowledged for his collaboration in the design and in the construction of the setup for the piezoelectric strain measurement.
This research has been partially supported by the Italian Ministry of Health, Grant Number RF-2011-02350464 (to G.C.) and by the European COST Action MP1202 (to M.L.).

Received: ((will be filled in by the editorial staff))
Revised: ((will be filled in by the editorial staff))
Published online: ((will be filled in by the editorial staff))

https://onlinelibrary.wiley.com/doi/10.1002/adhm.201600245[39]  M. Labardi, V. Likodimos, M. Allegrini, *Phys. Rev. B* **2000**, *61*, 14390.

[40]  S. F. Mendes, C. Costa, C. Caparros, V. Sencadas, S. Lanceros-Méndez, *J. Mater. Sci.* **2012**, *47*, 1378.

[41]  G. Ferrari-Toninelli, S. A. Bonini, D. Uberti, L. Buizza, P. Bettinsoli, P. L. Poliani, F. Facchetti, M. Memo, *Neuro-Oncology* **2010**, *12*, 1231.
24

**Figure 1.** Scanning electron microscopy images at different magnifications of P(VDF-TrFE) films (**a**: surface, **c** and **e**: cryosections) and of P(VDF-TrFE)/BTNP films (**b**: surface, **d** and **f**: cryosections).

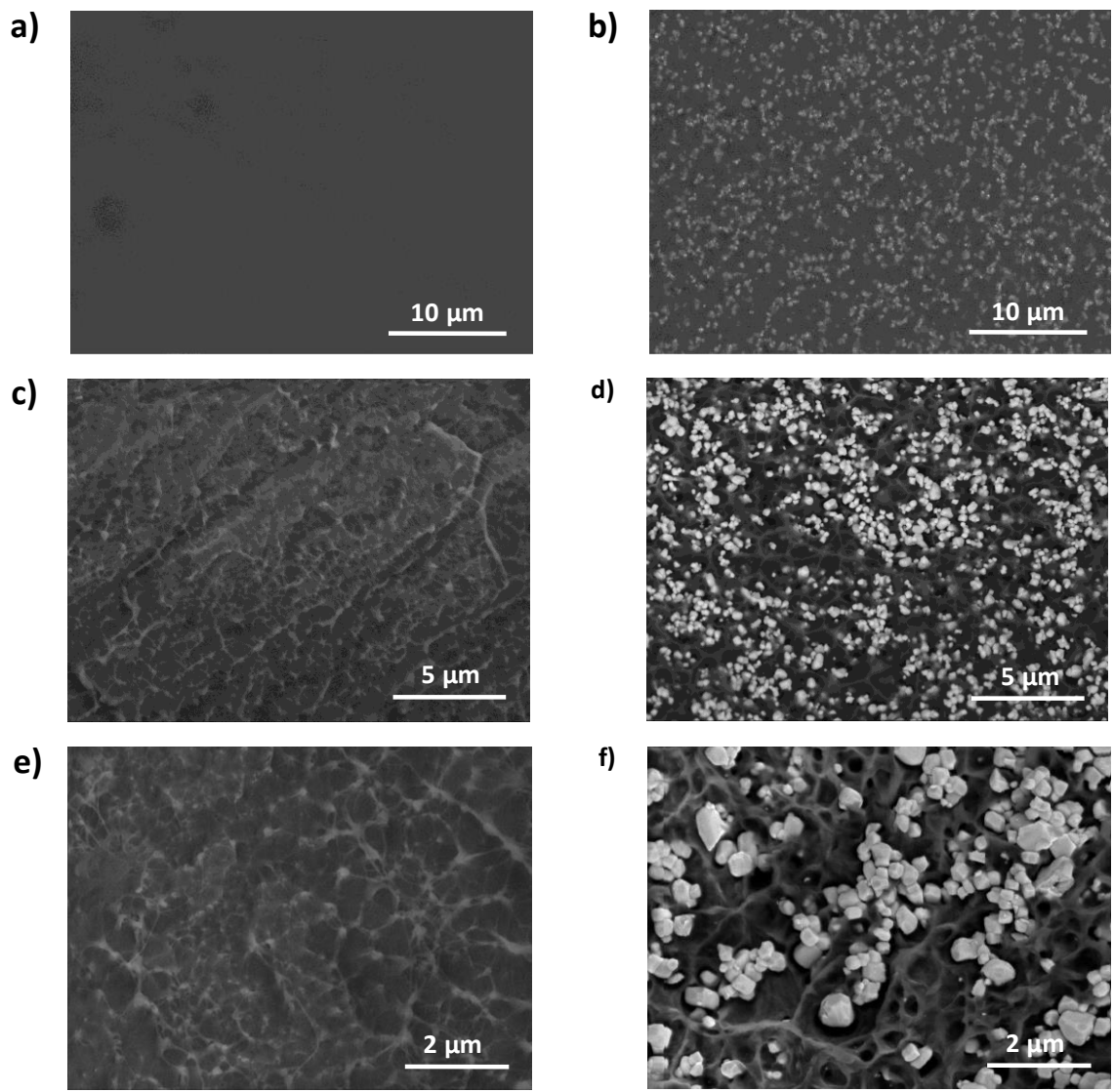

**Figure 2.** Atomic force microscopy images of P(VDF-TrFE) films (**a**) and P(VDF-TrFE)/BNTP films (**b**). Piezoresponse force microscopy maps of P(VDF-TrFE) films (**c**) and P(VDF-TrFE)/BNTP films (**d**).

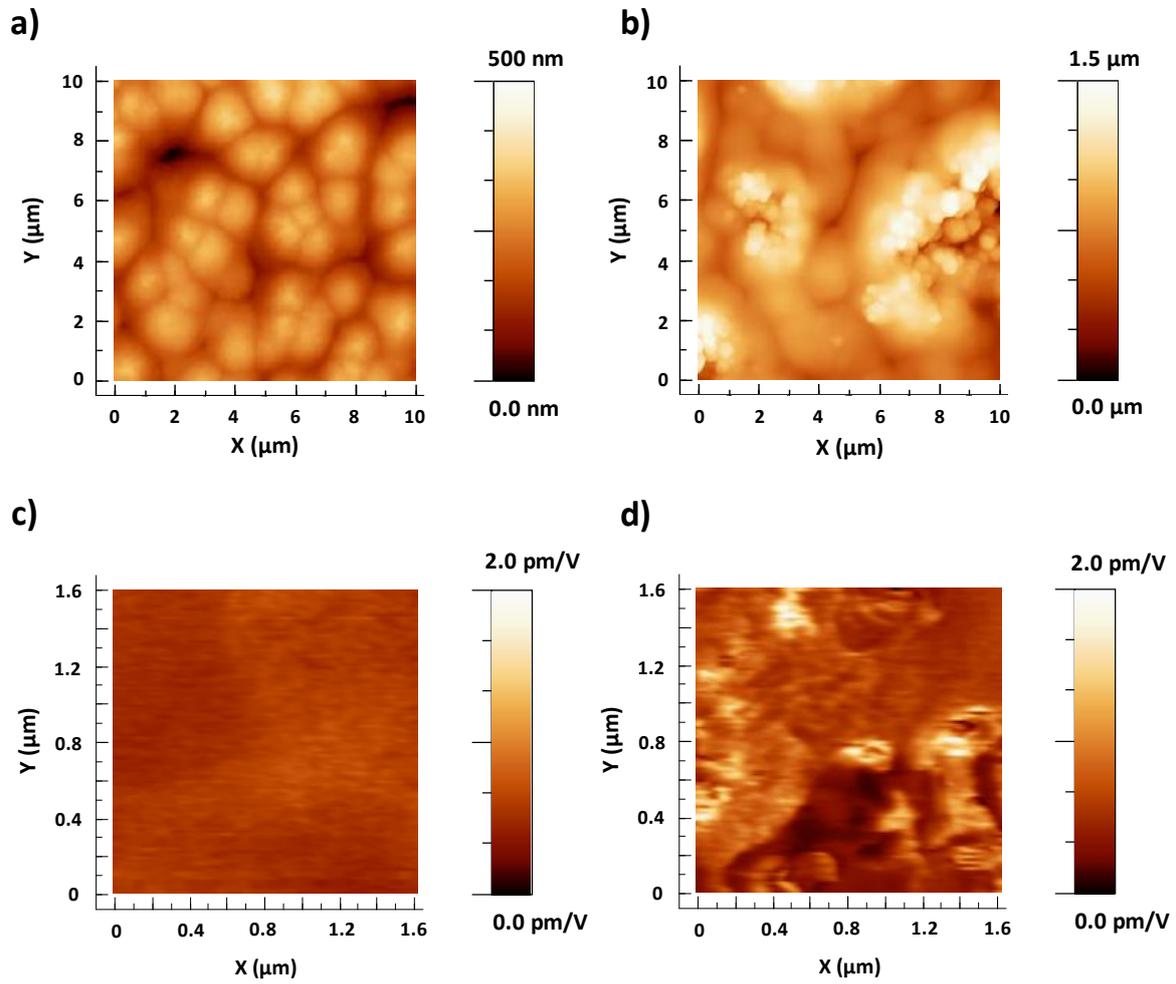

**Figure 3.** Differential scanning calorimetry thermograms (**a**), thermogravimetric analysis thermograms (**b**), and X-ray diffraction pattern (**c**) of P(VDF-TrFE) and P(VDF-TrFE)/BTNP films.

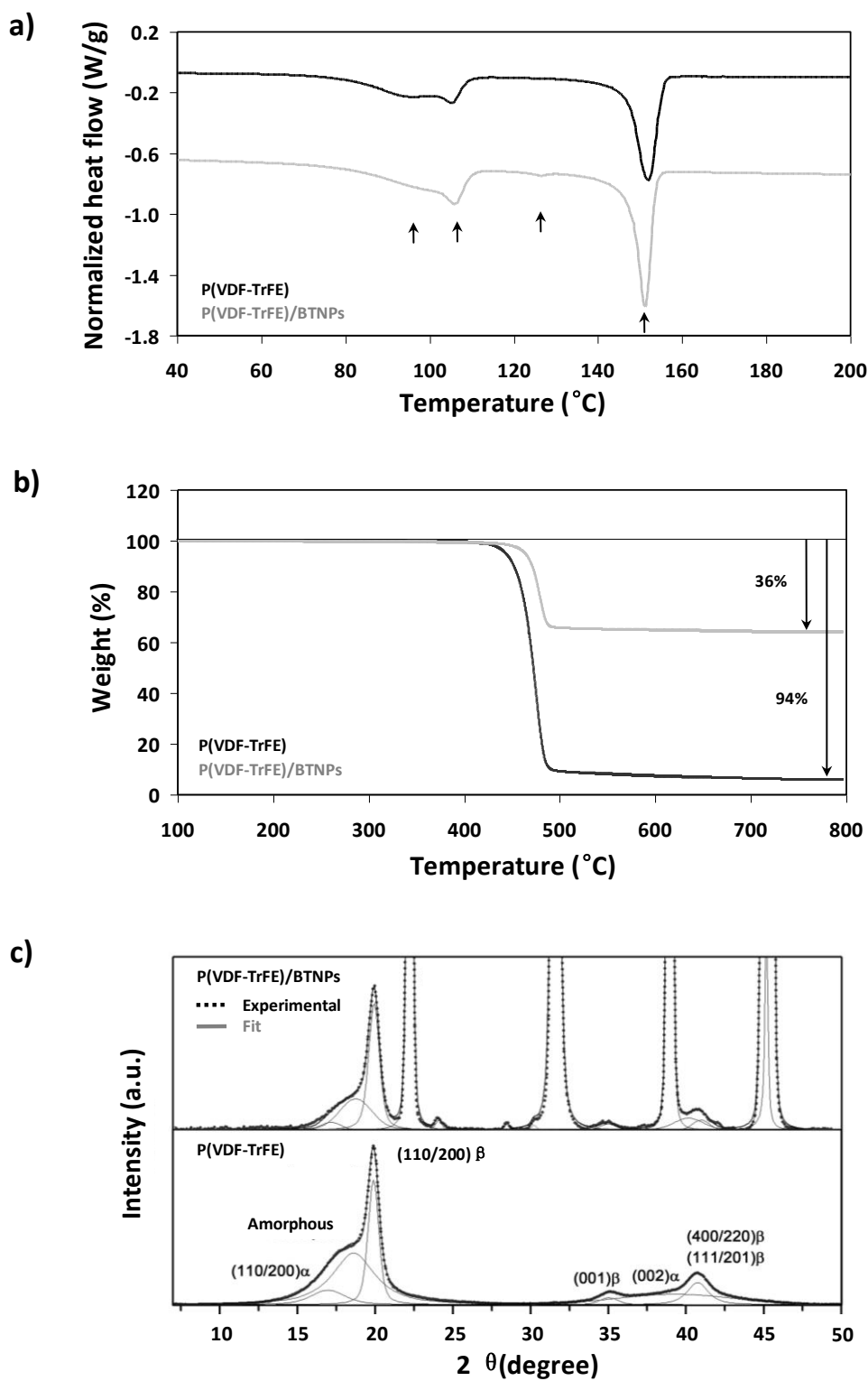

**Figure 4.** Extensometry of P(VDF-TrFE) and P(VDF-TrFE)/BTNP films. Representative stress-strain curves (**a**), Young's modulus (**b**), ultimate tensile strength (**c**), elongation at maximum load (**d**), and elongation at break (**e**). * $p < 0.05$.

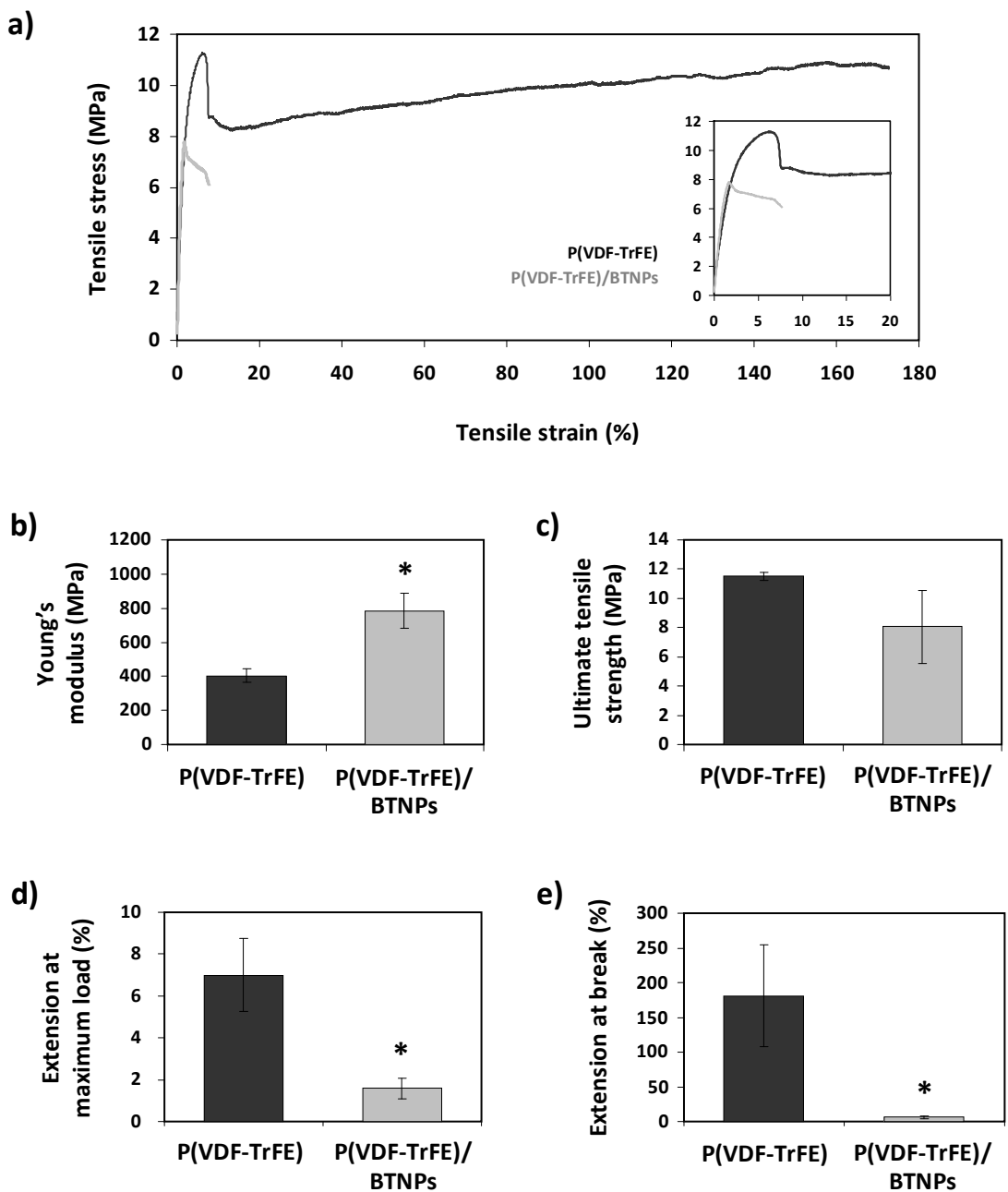

**Figure 5.** Fluorescence microscopy images of SH-SY5Y neuroblastoma cells following Live/Dead staining after 24 h of culture under proliferative conditions on P(VDF-TrFE) and P(VDF-TrFE)/BTNP films, and on Ibidi film as control. Live cells in green, dead cells in red (in this case no evidence of red staining), nuclei in blue (a). Cell density quantified with PicoGreen assay after 24 h and 72 h of cell culture under proliferative conditions on P(VDF-TrFE) and P(VDF-TrFE)/BTNP films, and on Ibidi film as control (b). *$p < 0.001$.

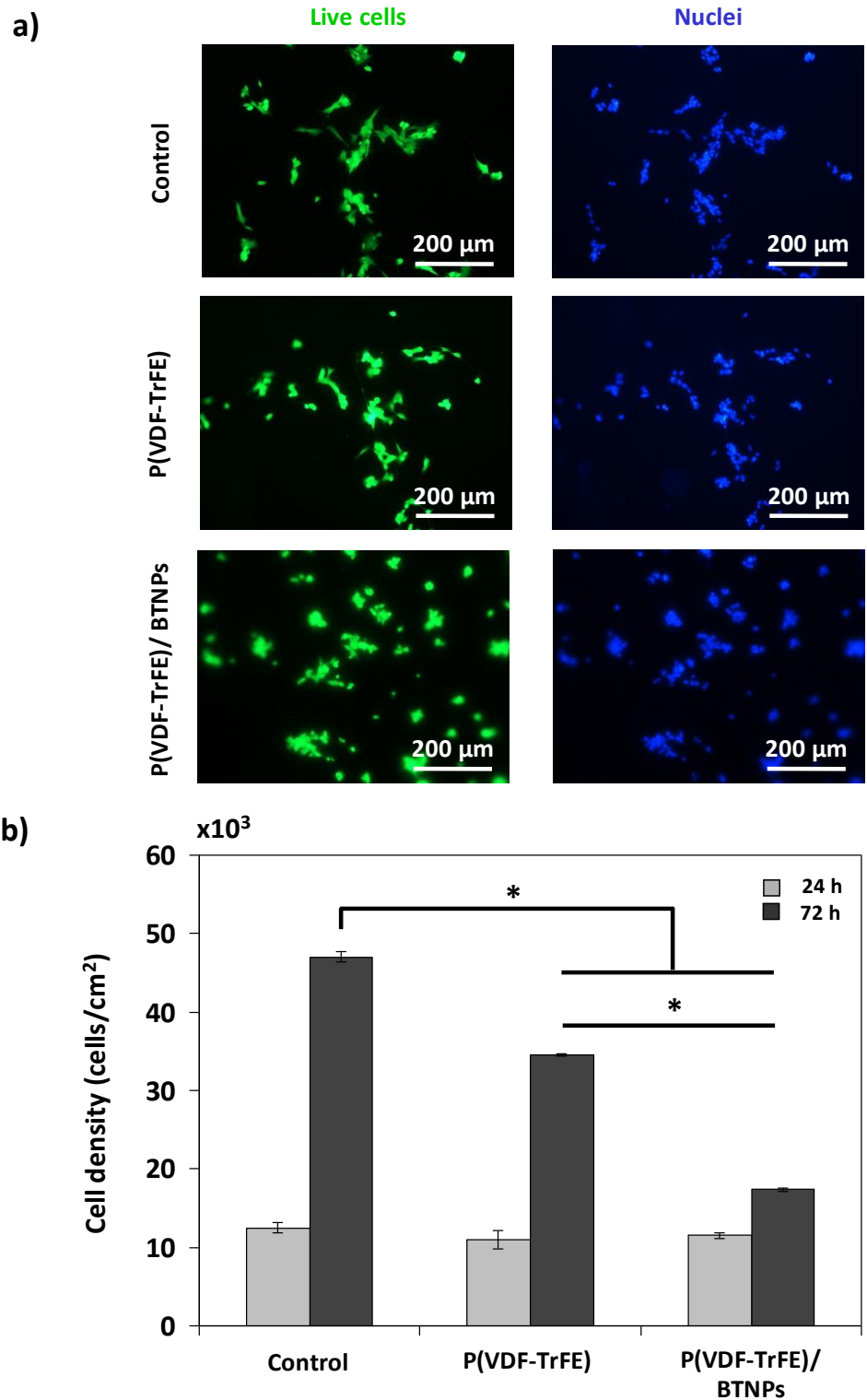

**Figure 6.** Calcium imaging analysis of SH-SY5Y neuroblastoma cells differentiated on P(VDF-TrFE) and P(VDF-TrFE)/BTNP films, and on Ibidi film as control, following US stimulation. Arrows indicate the time of US application. On the right column, a representative frame of the calcium imaging time-lapse (at $t = 50$ s).

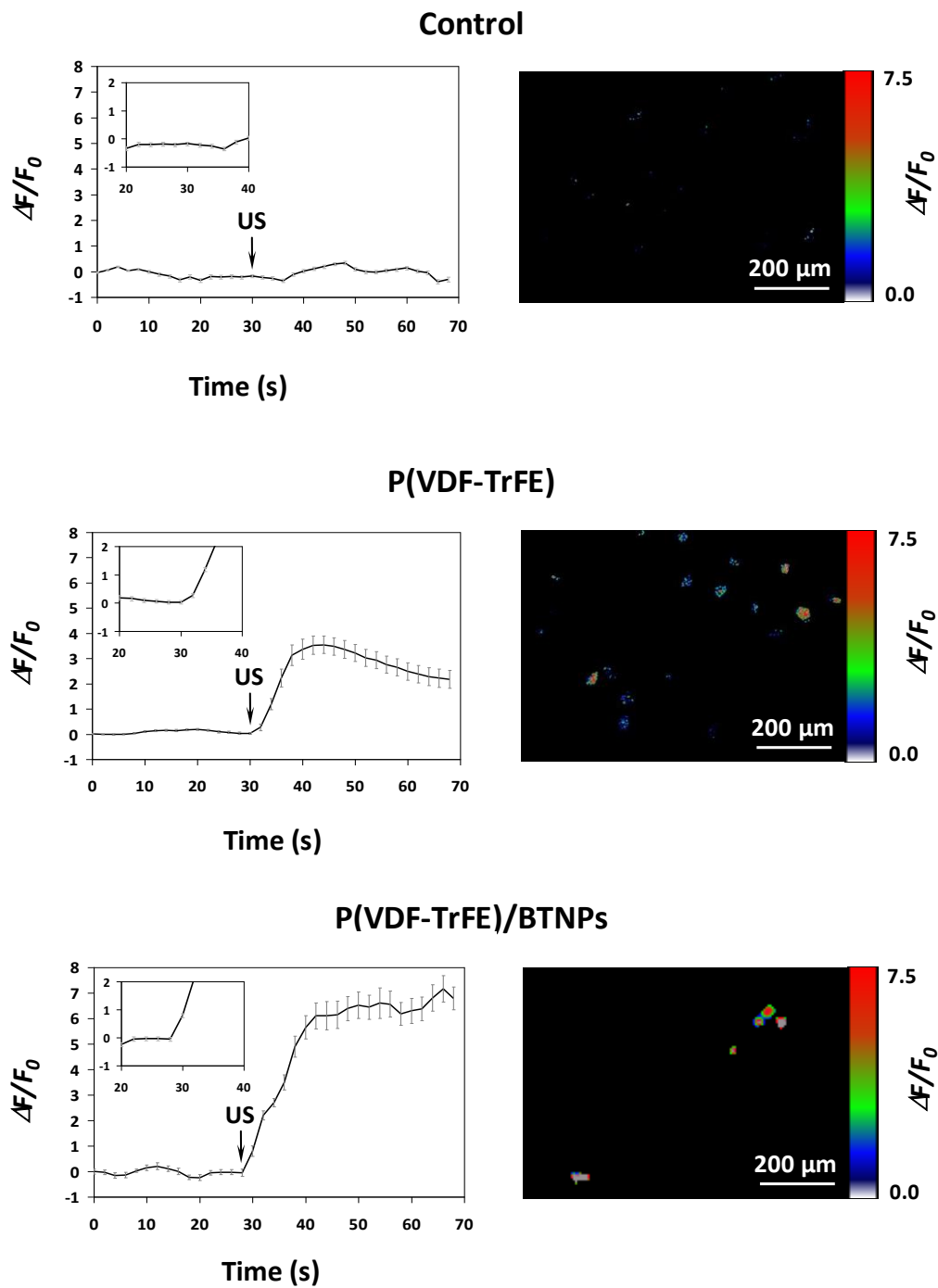

**Figure 7.** Confocal fluorescence microscopy images of SH-SY5Y neuroblastoma cells at the end of a differentiation period of 6 days on P(VDF-TrFE) and P(VDF-TrFE)/BTNP films, and on Ibidi film as control. Cells were either unexposed or exposed to chronic US stimulation. β3-tubulin is stained in green, nuclei in blue (a). Percentages of β3-tubulin positive cells (b). Neurite lengths are expressed as median values ± confidence interval at 95% (c). *$p < 0.05$.

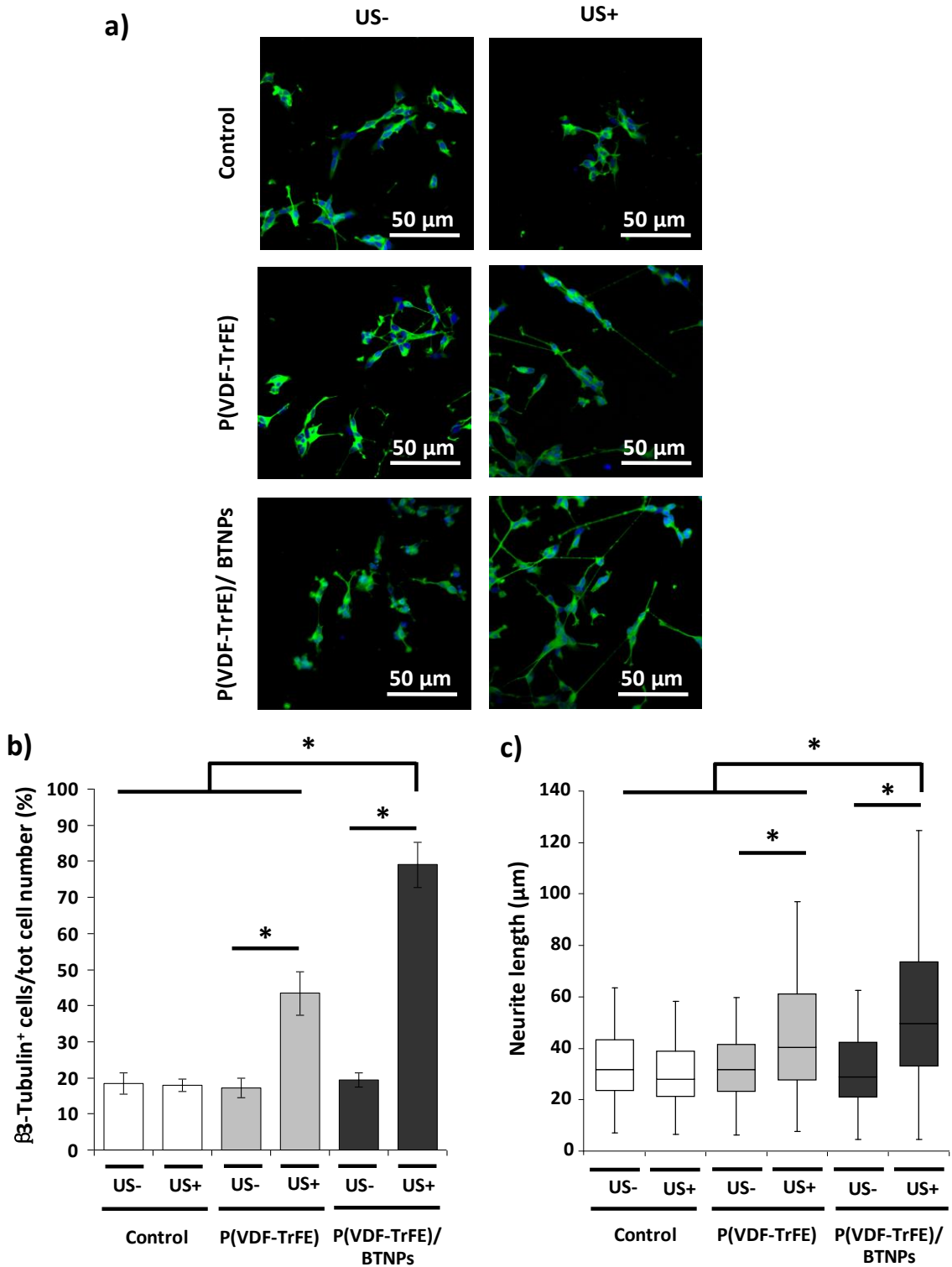

**Figure 8.** Scanning electron microscopy images of SH-SY5Y neuroblastoma cells at the end of a differentiation period of 6 days on P(VDF-TrFE) and P(VDF-TrFE)/BTNP films, and on Ibidi film as control. Cells were either unexposed or exposed to chronic US stimulation.

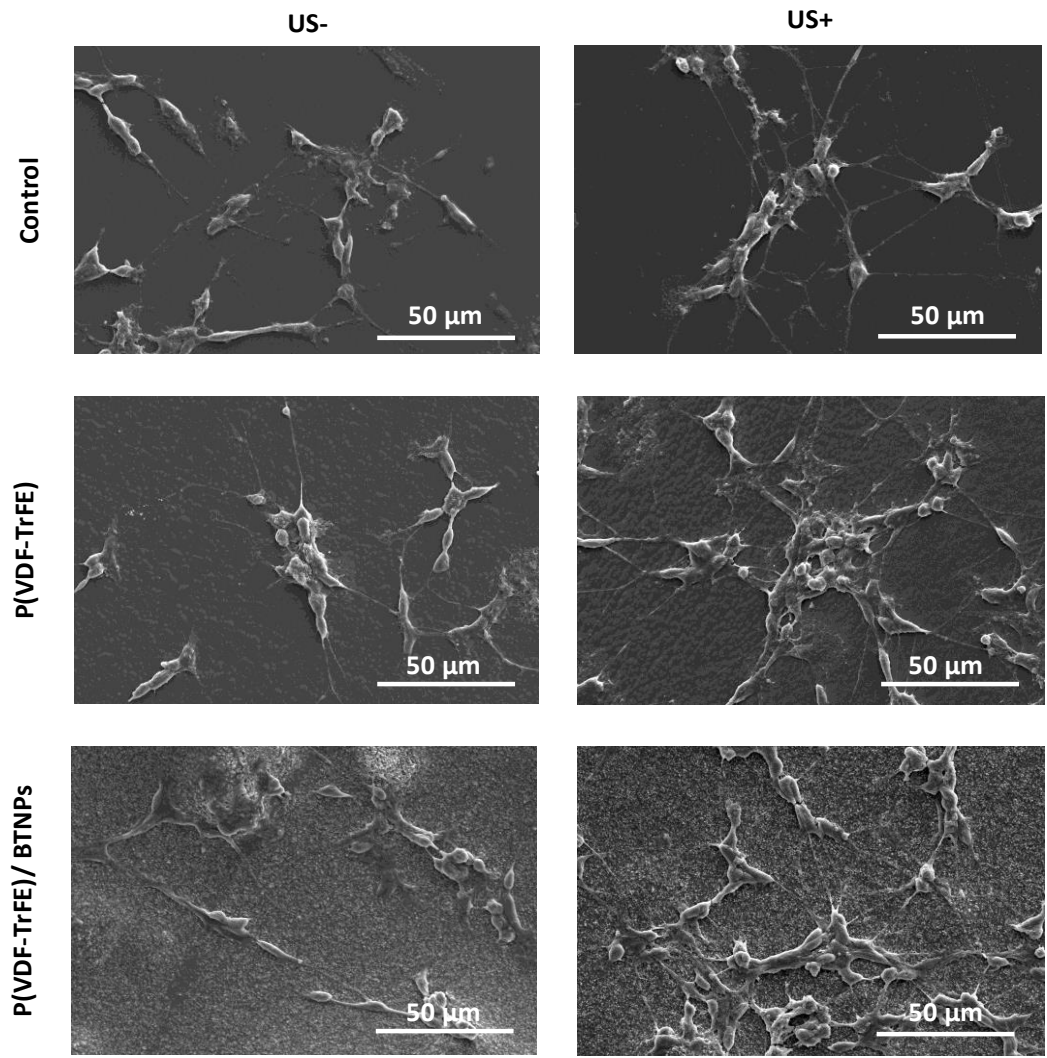

**Table 1.** Characterization of topography and piezoelectric response of P(VDF-TrFE) and P(VDF-TrFE)/BTNP films.

|  | PTFE (non-piezo) | P(VDF-TrFE) | P(VDF-TrFE)/BTNPs |
|---|---|---|---|
| RMS roughness (AFM) nm | 63 | 212 | 15.7 |
| Quality factor, $Q$ adimensional, ± 1% | 22.7 | 18.6 | 22.5 |
| $\varepsilon_d$ adimensional, ± 5% | 2.1 | 12.0 | 25.7 |
| Converse piezoelectric coefficient, $d_{31}$ pm/V ± 10% | 1.1 | 11.8 | 53.5 |
| Direct piezoelectric coefficient, $g_{31}$ m·V/N ± 10% | 0.06 | 0.11 | 0.24 |

**Table 2.** Characterization of bulk properties of P(VDF-TrFE) and P(VDF-TrFE)/BTNP films.

|  |  | P(VDF-TrFE) | P(VDF-TrFE)/BTNPs |
|---|---|---|---|
| Residual percentage weight (TGA) % ± 1 |  | 6 | 64 |
| Crystallinity (XRD) % ± 3.0 | α phase | 27.1 | 15.8 |
|  | β phase | 32.3 | 52.6 |
|  | total | 59.4 | 68.4 |
| *E* (extensometry) MPa |  | 404 ± 40 | 784 ± 103 |
| *UTS* (extensometry) MPa |  | 11.5 ± 0.3 | 8.1 ± 2.5 |
| *EML* (extensometry) % |  | 7 ± 2 | 2 ± 1 |
| *EB* (extensometry) % |  | 181 ± 73 | 6 ± 2 |

**P(VDF-TrFE) and P(VDF-TrFE)/BaTiO$_3$ nanoparticle composite films are used for SH-SY5Y stimulation.** Stimulation with ultrasounds of cells differentiated on composite films induces Ca$^{2+}$ transients by direct piezoelectric effect, determining an improvement of the neuronal phenotype. Suitability of P(VDF-TrFE)/BaTiO$_3$ NP films as active substrates in tissue engineering is proposed.

**Piezoelectric films**

G.G. Genchi*, L. Ceseracciu, A. Marino, M. Labardi, S. Marras, F. Pignatelli, L. Bruschini, V. Mattoli, G. Ciofani*

**P(VDF-TrFE)/BaTiO$_3$ nanoparticle composite films mediate piezoelectric stimulation and promote differentiation of SH-SY5Y neuroblastoma cells**

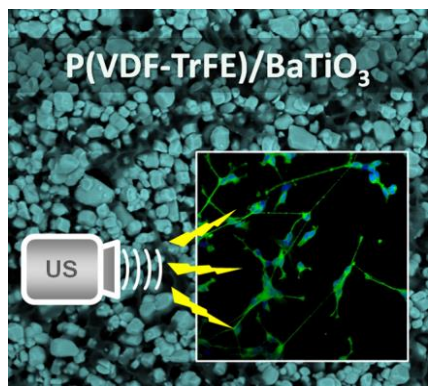

**Supporting information**

**P(VDF-TrFE)/BaTiO$_3$ nanoparticle composite films mediate piezoelectric stimulation and promote differentiation of SH-SY5Y neuroblastoma cells**

*Giada Graziana Genchi\*, Luca Ceseracciu, Attilio Marino, Massimiliano Labardi, Sergio Marras, Francesca Pignatelli, Luca Bruschini, Virgilio Mattoli, Gianni Ciofani\**

Dr. G.G. Genchi, A. Marino, Dr. F. Pignatelli, Dr. V. Mattoli, Prof. G. Ciofani
Istituto Italiano di Tecnologia, Center for Micro-BioRobotics @SSSA, Viale Rinaldo Piaggio 34, 56025 Pontedera (Pisa), Italy
A. Marino
Scuola Superiore Sant'Anna, The BioRobotics Institute, Viale Rinaldo Piaggio 34, 56025 Pontedera (Pisa), Italy
Dr. L. Ceseracciu
Istituto Italiano di Tecnologia, Nanophysics Department, Via Morego 30, 16163, Genova, Italy
Dr. S. Marras
Istituto Italiano di Tecnologia, Nanochemistry Department, Via Morego 30, 16163 Genova, Italy
Dr. M. Labardi
CNR-IPCF, Largo Pontecorvo 3, 56127 Pisa, Italy
Prof. L. Bruschini
University Hospital of Pisa, ENT Audiology and Phoniatry Unit, Via Paradisa 3, 56124 Pisa Italy
Prof. G. Ciofani
Politecnico di Torino, Department of Aerospace and Mechanical Engineering, Corso Duca degli Abruzzi 24, 10129 Torino, Italy

E-mail: giada.genchi@iit.it, gianni.ciofani@polito.it

Keywords: piezoelectric stimulation, neurons, P(VDF-TrFE), BaTiO$_3$ nanoparticles

**Figure S1.** X-ray diffraction pattern of P(VDF-TrFE)/BTNP films. Arrows indicate the peaks characteristic of tetragonal BaTiO$_3$, occurring at 44.98° and 45.37°.

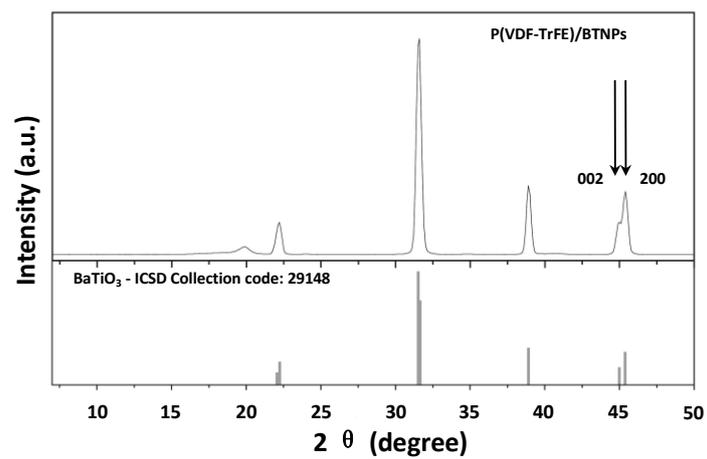

**Figure S2.** Schematization of the set up used for converse piezoelectric effect measurement.

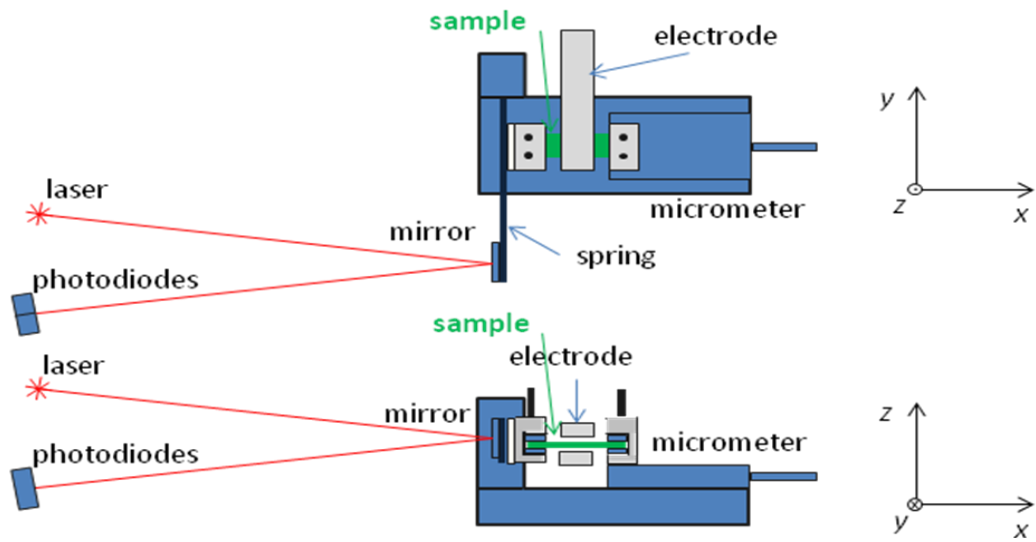